\documentclass[amsmath,amsfonts,prl,aps,twocolumn,twoside,superscriptaddress]{revtex4}

\newcommand{\be}{\begin{equation}}
\newcommand{\ee}{\end{equation}}
\newcommand{\ba}{\begin{array}}
\newcommand{\ea}{\end{array}}
\newcommand{\bea}{\begin{eqnarray}}
\newcommand{\eea}{\end{eqnarray}}
\newcommand{\la}{\langle}
\newcommand{\ra}{\rangle}
\newcommand{\nn}{\nonumber}
\newcommand{\trace}{\mathop{\mathrm{Tr}}\nolimits}
\newcommand{\RR}{\mathbb{R}}
\newcommand{\CC}{\mathbb{C}}
\newcommand{\calH}{{\cal H }}
\newcommand{\ket}[1]{| #1 \rangle}

\newtheorem{lemma}{Lemma}
\newtheorem{definition}{Definition}
\newtheorem{example}{Example}

\begin{document}
\title{Unextendible maximally entangled bases}
\author{Sergei Bravyi}\email{sbravyi@us.ibm.com}
\author{John A. Smolin}\email{smolin@watson.ibm.com}
\affiliation{IBM T.J. Watson Research Center, Yorktown Heights, NY 10598, USA}

\date{\today}

\begin{abstract}
We introduce the notion of the unextendible maximally entangled basis
(UMEB), a set of orthonormal maximally entangled states in
$\CC^d\otimes \CC^d$ consisting of fewer that $d^2$ vectors which have
no additional maximally entangled vectors orthogonal to all of them.
We prove that UMEBs don't not exist for $d=2$ and give an explicit
constructions for a 6-member UMEB with $d=3$ and a 12-member UMEB with
$d=4$.
\end{abstract}
\maketitle

Ever since Einstein, Podolsky, and Rosen (EPR) demonstrated the
necessarily nonlocal nature of quantum mechanics \cite{EPR} it has
been recognized that the difference between factorizable and product
(nonfactorizable) quantum states is fundamental to understanding the
deepest implications of quantum mechanics to information theory and
even to the nature of reality \cite{BELL}.  The centrality of this
distinction made it all the more surprising when it was found that
there are sets of product states which nevertheless display a form of
nonlocality \cite{UPB1,UPB2}.  There it was shown that there are sets
of orthogonal product vectors of a tensor product Hilbert space $\CC^n
\otimes \CC^m$ ($n,m>2$) such that there are no further product states orthogonal
to every state in the set, even though the size of the set is smaller
than $nm$.  These unextendible product bases (UPBs) are not
distinguishable by local measurements with classical communication.  It
was also shown that the space complementary to a UPB contains bound
entanglement \cite{bound}.  Later, after generalizing the concept such
that the states in the set need not be product states, Duan used
unextendible bases to show superactivation of the zero-error capacity
of a quantum channel \cite{Duan0}.

Here, we generalize the notion of the UPB to the unextendible entangled basis.
This is a set of bipartite pure states $|\Psi_i\rangle$
each of which has entanglement $\alpha$ but whose complementary space is
non-empty and contains no states of entanglement $\alpha$.  We will restrict
our attention here to the unextendible maximally entangled bases (UMEBs) of
$\CC^d\otimes \CC^d$ ($\alpha=\log d$).  We show that there are no
UMEBs for $d=2$ and give a constructive examples
of a 6-member UMEB with $d=3$ and a 12-member UMEB for $d=4$.

\begin{definition}
A set of states $\{|\Psi_a\ra\in\CC^d\otimes \CC^d\}$, $a=1,\ldots,n$
is called an UMEB iff\\ (i) All states $|\Psi_a\ra$ are maximally
entangled;\\ (ii) $\la \Psi_a|\Psi_b\ra=\delta_{a,b}$;\\ (iii) If $\la
\Psi_a|\Psi\ra=0$ for all $a=1,\ldots,n$ then $|\Psi\ra$ cannot be
maximally entangled.
\end{definition}

It is convenient to represent a basis vector $\Psi_a$ of UMEB by a
linear operator $U_a$ acting on $\CC^d$ such that
\begin{equation}
|\Psi_a\ra=(I\otimes U_a)\, |\Phi\ra, \quad |\Phi\ra =\frac1{\sqrt{d}}\, \sum_{j=1}^d |j,j\ra.
\label{IotimesUPhi}
\end{equation}
The conditions (i-iii) can now be rephrased as

{\em
\noindent
(i) Operators $U_1,\ldots,U_n$ are unitary;\\
(ii) $\trace{(U_a^\dag\, U_b)}=d\, \delta_{a,b}$;\\
(iii) If $\trace{(U_a^\dag\, U)}=0$ for all $a=1,\ldots,n$ then $U$ cannot be unitary.
}

The smallest nontrivial bipartite Hilbert space is $\CC^2\otimes\CC^2$.  We
first show that such a space does not admit a UMEB:
\begin{lemma}
UMEBs do not exist in $\CC^2\otimes \CC^2$.
\end{lemma}
{\bf Proof:}
Clearly, if $\{U_1,\ldots,U_n\}$ is UMEB and $V$ is a unitary operator
then $\{VU_1,\ldots,VU_n\}$ is UMEB and $\{U_1V,\ldots,U_nV\}$ is UMEB. Thus without loss
of generality
\[
U_1=I, \quad U_2 =\sigma_z.
\]
Since we can extend $\{U_1,U_2\}$, there must be at least three basis vectors. We can always
write
\[
U_3=\alpha \, \sigma_x + \beta\, \sigma_y, \quad |\alpha|^2 + |\beta|^2=1.
\]
But now we can complete the basis by
$U_4=\beta^*\, \sigma_x - \alpha^*\, \sigma_y$.
\begin{flushright}
$\Box$
\end{flushright}
More generally, UMEB in $\CC^d\otimes \CC^d$ cannot have $d^2-1$ basis vectors since
\[
|\Psi\ra\la \Psi|:=I-\sum_{a=1}^{d^2-1} |\Psi_a\ra\la \Psi_a|
\]
defines a maximally entangled state orthogonal to all $d^2-1$ basis vectors.

\begin{example}UMEB in $\CC^3\otimes \CC^3$ (ICOSAHEDRON)\end{example}
Let us now construct an explicit example of UMEB in $\CC^3\otimes \CC^3$
with $n=6$ basis vectors.  Consider the following 6 vectors in $\CC^3$, the
diagonals of the icosahedron:
\begin{eqnarray}
\label{psis}
\nonumber\ket{\psi_{1,2}}=\frac{1}{N}\left(\ket{0} \pm \phi\ket{1}\right)\\
\ket{\psi_{3,4}}=\frac{1}{N}\left(\ket{1} \pm \phi\ket{2}\right)\\
\nonumber\ket{\psi_{5,6}}=\frac{1}{N}\left(\ket{2} \pm \phi\ket{0}\right)
\end{eqnarray}
with $\phi=(1+\sqrt{5})/2$, the golden ratio, and $N=\sqrt{1+\phi^2}$
as normalization.  Note that  Eq.~(\ref{psis}) defines an
equiangular set of vectors, that is,
\be
\label{angles}
|\langle \psi_j|\psi_k\rangle|^2 = \frac15 \quad \mbox{for all $j\ne k$}.
\ee
Define unitary operators
\begin{equation}
U_j= I-(1-e^{i \theta}) |\psi_j\rangle\!\langle\psi_j|, \quad j=1,\ldots,6.
\end{equation}
Using Eq.~(\ref{angles}) one can check that
these unitaries are pairwise orthogonal,  $\trace(U^\dag_a U_b)=3 \delta_{a,b}$,
provided that $\cos{\theta}=-7/8$.

Since the vectors $|\psi_j\ra$ are real, the operators $U_j$ are represented by
symmetric matrices in the computational basis $|0\rangle,|1\rangle,|2\rangle$.
It means that the
corresponding states
\be
|U_j\ra\equiv (I\otimes U_j) \, |\Phi\rangle, \quad j=1,\ldots,6
\ee
 belong to the symmetric subspace
$\calH_{\rm sym}$ of two qutrits. Moreover, the states $|U_j\ra$, $j=1,\ldots,6$ form an
orthonormal basis of the symmetric subspace since  $\dim{\calH_{\rm sym}}=6$.
By construction, the subspace orthogonal to $|U_1\rangle,\ldots,|U_6\rangle$
coincides with the antisymmetric subspace $\calH_{\rm asym}$ of two qutrits.
Suppose there exists a maximally entangled state
\be
|U\ra = (I\otimes U)\, |\Phi\rangle \in \calH_{\rm asym}, \quad U^\dag U=I.
\ee
This is possible only if $U$ is skew-symmetric matrix, $U^T=-U$.
But a skew-symmetric matrix of odd dimension cannot be unitary since it has zero determinant.
Thus $\calH_{\rm asym}$ contains no maximally entangled states.

\begin{example}UMEB in $\CC^4\otimes \CC^4$ (TILES)  \end{example}
We now construct a UMEB in $\CC^4\otimes \CC^4$ with $n=12$ basis
vectors.  This UMEB has an interesting connection to the UPB ``TILES''
defined in \cite{UPB1,UPB2}.  We shall identify $\CC^4$ with the
Hilbert space of two qubits and all operators $U_a$ will be written in
terms of the Pauli matrices $\sigma_\alpha$, where $\alpha=x,y,z$.
For any vector $\vec{u}\in \RR^3$ we shall denote
$\sigma(\vec{u})=u^x\, \sigma_x + u^y\, \sigma_y+u^z\, \sigma_z$.
First of all, choose \be \{U_1,\ldots,U_5\} = \{
\sigma(\vec{u_a})\otimes \sigma(\vec{v_a}),\quad a=1,\ldots,5 \}, \ee
where vectors $\{|\vec{u}_a\ra\otimes |\vec{v}_a\ra\}$ are the members
of the TILES UPB. Explicitly, \bea U_1
&=&\frac{1}{\sqrt{2}} \sigma_x\otimes (\sigma_x-\sigma_y), \nn \\ U_2
&=&\frac{1}{\sqrt{2}}(\sigma_x-\sigma_y)\otimes \sigma_z, \nn \\ U_3
&=&\frac{1}{\sqrt{2}}\sigma_z\otimes (-\sigma_y+\sigma_z), \nn \\ U_4
&=&\frac{1}{\sqrt{2}}(-\sigma_y+\sigma_z)\otimes \sigma_x, \nn \\ U_5
&=&\frac1{3}(\sigma_x+\sigma_y+\sigma_z)\otimes
(\sigma_x+\sigma_y+\sigma_z). \nn \eea Using an identity
$\trace{(\sigma(\vec{u})\, \sigma(\vec{v}))}=2(\vec{u},\vec{v})$ and
the fact that $\{|\vec{u}_a\ra\otimes |\vec{v}_a\ra\}$ form UPB we
infer that is not possible to extend $\{U_1,\ldots,U_5\}$ by a unitary
operator $U=\sigma(\vec{u})\otimes \sigma(\vec{v})$. In order to get
UMEB, we choose \be \{U_6,\ldots,U_{12}\}=\{I\otimes I,\; I\otimes
\sigma_\alpha,\; \sigma_\beta\otimes I\}.  \ee We get twelve unitary
operators. Suppose one can extend the basis by $U$.  Orthogonality to
$U_6,\ldots,U_{12}$ implies that \be
\label{U}
U=\sum_{\alpha,\beta=x,y,z} A_{\alpha,\beta}\, \sigma_\alpha \otimes \sigma_\beta.
\ee
Orthogonality to $U_1,\ldots,U_5$ implies that
\be
\label{A}
A=\left( \ba{ccc} a & a & b \\ d & g & b \\ d & c & c \\ \ea \right), \quad g=-2(a+b+c+d),
\ee
for some complex numbers $a,b,c,d$.

Assume that the operator $U$ defined in Eqs.~(\ref{U},\ref{A}) is unitary. 
Computing $\trace{ (UU^\dag A\otimes B)}$ for
$A,B\in \{I,\sigma_x,\sigma_z\}$ one gets after simple algebra
\be
\label{aux1}
(g+a)\bar{b}=0, \quad (g+b)\bar{c}=0, \quad (g+c)\bar{d}=0, \quad (g+d)\bar{a}=0,
\ee
\be
\label{norm1}
|a|^2 + |b|^2+|c|^2+|d|^2 + |g|^2/2=1/2.
\ee
Let $K:=\sigma_y\otimes (\sigma_x+\sigma_z) + (\sigma_x+\sigma_z)\otimes \sigma_y - \sigma_y\otimes \sigma_y$.
Using the identity $\trace{(UU^\dag K)}=0$ one gets after simple algebra 
\be
\label{norm2}
|a+b+c+d|^2=|a|^2 + |b|^2 + |c|^2 +|d|^2.
\ee
Combining Eqs.~(\ref{norm1},\ref{norm2}) with $g=-2(a+b+c+d)$ we infer that $g\ne 0$. 
Then one can easily check that Eq.~(\ref{aux1}) has the only solution $a=b=c=d=0$
which contradicts $g\ne 0$. Thus $U$ cannot be unitary.

{\em Applications:}
UMEBs can be used to construct examples of states for which 1-copy
entanglement of assistance (EoA) \cite{EoA} is strictly smaller than the
asymptotic EoA. Indeed, a state proportional to the projector onto the UMEB's
complementary subspace
\begin{equation}
\label{rhocomplement}
\rho^\perp=\frac1{d^2-n} \left( I-\sum_{a=1}^n |\Psi_a\ra\la \Psi_a|\right)
\end{equation}
has no maximally entangled states in its range. Therefore the EoA of
$\rho^\perp$ satisfies $E_A(\rho^\perp)<\log{d}$. On the other hand, it was shown
in \cite{SVW} that for all
$\rho$
\[
E_A^\infty(\rho)=\lim_{k\to \infty} \frac1k E_A(\rho^{\otimes k}) = \min{\{S(\rho_1),S(\rho_2)\}},
\]
which gives $\log_2 d$ for any state $\rho^\perp$ of the form
(\ref{rhocomplement}).  The ICOSAHEDRON UMEB has as its complementary
space the anti-symmetric subspace of $\CC^3\otimes\CC^3$, which
contains no states of Schmidt rank higher than 2. Thus, the EoA of
that state is at most $1$ while the asymptotic EoA is $\log_2 3$.

Another property of a UMEBs is that they can be used to find quantum channels
that are unital but not convex mixtures of unitary operations.  That such
channels exist, in contrast to the classical analogue where
there are no doubly-stochastic channels which are not convex combinations of
permutations, was shown by Landau and Streater \cite{Streater}.  We now
show that the state $\rho^\perp$ is the
Choi-Jamio\l{}kowski matrix \cite{jamiolkowski} associated with such a channel.

For a matrix $\rho_{AB}$ on a bipartite $d\times d$ system $\calH_A \otimes \calH_B$ to be a
Choi-Jamio\l{}kowski matrix, it must have $\trace_B
\rho_{AB}=\frac{I}{d}$ and for the corresponding channel to to be
unital $\trace_A \rho_{AB}=\frac{I}{d}$.  Both of these a satisfied
by $\rho^\perp$ (every term in the sum has
$\trace_A \rho = \trace_B \rho \propto I$).  Finally, the
Choi-Jamio\l{}kowski matrix associated with a mixture of unitaries
is always of the form
\[
\rho_{\rm CJ}=\sum_a (I\otimes U_a) |\Phi\ra\!\la \Phi| (I\otimes U_a)^\dag
\]
{\em i.e.} it is a mixture of maximally entangled states.  Clearly
$\rho^\perp$ cannot have this form as by construction it has no
maximally entangled states in its support.

Note that $\rho^\perp$ for the ICOSAHEDRON UMEB, the projector onto
the antisymmetric subspace, is the same as the unital but unitary
example found in \cite{Streater}, except that in that construction there
are some different but unimportant phases.

{\em Discussion:}
We have shown that there exist no UMEBs in $\CC^2\otimes\CC^2$ and
have given explicit examples in $\CC^3\otimes \CC^3$ and $\CC^4\otimes
\CC^4$.  It is also possible to generalize the concept of UMEBS
to the case where the bipartite spaces are of unequal size.  One has
only to restrict the sum in Eq. (\ref{IotimesUPhi}) the the smaller of
the two dimensions.  We do not know of any such nonsquare UMEBS exist,
but if they do they might have interestingly different properties.  For
instance, the relation the complementary space to unital channels would
be broken, since it is no longer true that ever term in (\ref{rhocomplement})
has $\trace_A \rho = \trace_B \rho \propto I$.  Examples in
$\CC^2 \otimes \CC^3$ or $\CC^2\otimes\CC^4$ would be particularly interesting,
as the total dimensions 6 and 8 are smaller than the 9-dimensional example of
$\CC^3\otimes\CC^3$.

There remain many other open questions about UMEBS.  Are there UMEBS
for all $d>2$?  What are the the minimum and maximum numbers of states
in a UMEB as a function of dimension.  We hope that unextendible
entangled bases prove as useful to quantum information theory as UPBs
did before them.

{\em Acknowledgments:} This work has been supported by DARPA QUEST
program under contract no. HR0011-09-C-0047.  We would like to thank
Graeme Smith for helpful discussions and pointing out the relation of
UMEBS to unital channels that are not decomposable as mixtures of
unitaries.


\begin{thebibliography}{99}

\bibitem{EPR} A. Einstein, B. Podolsky, N. Rosen, Phys. Rev. {\bf 47}, 777 (1935).

\bibitem{BELL} J.S. Bell, Physics (N.Y.) {\bf 1}, 195 (1964).

\bibitem{UPB1}C.H. Bennett, D.P. Divincenzo, T. Mor, P.W. SHor,
  J.A. Smolin and B.M. Terhal, ``Unextendible Product Bases and Bound
  Entanglement,'' Phys. Rev. Lett. {\bf 82}, 5385 (1999).


\bibitem{UPB2}D.P. Divincenzo, T. Mor, P.W. Shor, J.A. Smolin and B.M. Terhal,
``Unextendible Product Bases, Uncompletable Product Bases and Bound
Entanglement,'' Commun. Math. Phys. 238, 379-410 (2003).

\bibitem{bound} Bound entanglement was defined in P. Horodecki, Phys. Lett. A
{\bf 232}, 333 (1997).


\bibitem{Duan0} R. Duan, ``Super-Activation of Zero-Error Capacity of
Noisy Quantum Channels,'' arxiv:0906.2527.

\bibitem{EoA}D.P. DiVincenzo, C.A. Fuchs, H. Mabuchi, J.A. Smolin, A.V.
Thapliyal and A. Uhlmann, ``Entanglement of Assistance,'' in: Proc.
{\em Quantum Computing and Quantum Communications: First NASA Intl. Conf.,
Palm Springs, 1998}, Springer LNCS 1509, pp. 247-257, Heidelberg, 1999.

\bibitem{SVW}
J.A. Smolin, F. Verstraete and A. Winter, ``Entanglement of assistance and multipartite state distillation,'' Phys. Rev. A {\bf 72}, 052317 (2005), quant-ph/0505038.

\bibitem{Streater} L. J. Landau, R. F. Streater, ``On Birkhoff's theorem for
    doubly stochastic completely positive maps of matrix algebras,''
    Lin. Alg. Appl., vol. 193, pp. 107-127, 1993.

\bibitem{jamiolkowski} M.-D. Choi, ``Completely positive linear maps
  on complex matrices,'' Linear Algebra Appl. {\bf 10}, 285-290
  (1975). A. Jamio\l{}kowski, ``Linear transformations which preserve
  trace and positive semidefiniteness of operators,''
  Rep. Math. Phys., vol. 3, pp. 275-278, 1972.

\end{thebibliography}
\end{document}